\documentclass[12pt]{article}
\usepackage{graphicx}
\usepackage{color}
\usepackage{authblk}

\def\hybrid{\topmargin 0pt      \oddsidemargin 0pt
        \headheight 0pt \headsep 0pt
       \voffset-1cm
        \textwidth 6.25in       
       \textheight 9.5in       
        \marginparwidth 0.0in
        \parskip 5pt plus 1pt   \jot = 1.5ex}
\catcode`\@=11
\def\marginnote#1{}

\newcount\hour
\newcount\minute
\newtoks\amorpm
\hour=\time\divide\hour by60
\minute=\time{\multiply\hour by60 \global\advance\minute by-\hour}
\edef\standardtime{{\ifnum\hour<12 \global\amorpm={am}%
        \else\global\amorpm={pm}\advance\hour by-12 \fi
        \ifnum\hour=0 \hour=12 \fi
        \number\hour:\ifnum\minute<10 0\fi\number\minute\the\amorpm}}
\edef\militarytime{\number\hour:\ifnum\minute<10 0\fi\number\minute}

\def\draftlabel#1{{\@bsphack\if@filesw {\let\thepage\relax
   \xdef\@gtempa{\write\@auxout{\string
      \newlabel{#1}{{\@currentlabel}{\thepage}}}}}\@gtempa
   \if@nobreak \ifvmode\nobreak\fi\fi\fi\@esphack}
        \gdef\@eqnlabel{#1}}
\def\@eqnlabel{}
\def\@vacuum{}
\def\draftmarginnote#1{\marginpar{\raggedright\scriptsize\tt#1}}

\def\draftlabel#1{{\@bsphack\if@filesw {\let\thepage\relax
   \xdef\@gtempa{\write\@auxout{\string
      \newlabel{#1}{{\@currentlabel}{\thepage}}}}}\@gtempa
   \if@nobreak \ifvmode\nobreak\fi\fi\fi\@esphack}
        \gdef\@eqnlabel{#1}}
\def\@eqnlabel{}
\def\@vacuum{}
\def\draftmarginnote#1{\marginpar{\raggedright\scriptsize\tt#1}}

\def\draft{\oddsidemargin -.5truein
        \def\@oddfoot{\sl preliminary draft \hfil
        \rm\thepage\hfil\sl\today\quad\militarytime}
        \let\@evenfoot\@oddfoot \overfullrule 3pt
        \let\label=\draftlabel
        \let\marginnote=\draftmarginnote
   \def\@eqnnum{(\theequation)\rlap{\kern\marginparsep\tt\@eqnlabel}%
\global\let\@eqnlabel\@vacuum}  }

\def\numberbysection{\@addtoreset{equation}{section}
        \def\theequation{\thesection.\arabic{equation}}}

\def\underline#1{\relax\ifmmode\@@underline#1\else
        $\@@underline{\hbox{#1}}$\relax\fi}

\def\titlepage{\@restonecolfalse\if@twocolumn\@restonecoltrue\onecolumn
     \else \newpage \fi \thispagestyle{empty}\c@page\z@
        \def\thefootnote{\fnsymbol{footnote}} }

\def\endtitlepage{\if@restonecol\twocolumn \else  \fi
        \def\thefootnote{\arabic{footnote}}
        \setcounter{footnote}{0}}  
\relax

\numberbysection
\hybrid

\newfont{\Bbb}{msbm10 scaled 1\@ptsize00}
\newfont{\Bbbb}{msbm7 scaled 1\@ptsize00}

\newcommand{\DDD}{\raise-1pt\hbox{$\mbox{\Bbbb D}$}}

\newcommand{\UUU}{\raise-1pt\hbox{$\mbox{\Bbbb U}$}}

\newcommand{\z}{\raise-1pt\hbox{$\mbox{\Bbbb Z}$}}

\def\res{\mathop{\hbox{res}}\limits}

\def\beq{\begin{equation}}
\def\eeq{\end{equation}}
\def\p{\partial}

\begin{document}

\begin{titlepage}

\title{Elliptic solutions to matrix KP hierarchy and spin generalization of elliptic
Calogero-Moser model}

\vspace{0.5cm}

\author[1,2]{V.~Prokofev\thanks{vadim.prokofev@phystech.edu }}
\author[2,3,4]{
 A.~Zabrodin\thanks{ zabrodin@itep.ru}}
 \affil[1]{Moscow Institute of Physics and Technology, Dolgoprudny, Institutsky per., 9,
Moscow region, 141700, Russia}
 \affil[2]{
Skolkovo Institute of Science and Technology, 143026 Moscow, Russian Federation
}
\affil[3]{ National Research University Higher School of Economics,
20 Myasnitskaya Ulitsa, Moscow 101000, Russian Federation}
\affil[4]{Institute of Biochemical Physics, 119334, Kosygina str. 4, Moscow, Russian Federation
}


\date{March 2021}
\maketitle


\begin{abstract}

We consider solutions of the matrix KP hierarchy that are elliptic 
functions of the first hierarchical time $t_1=x$. It is known that
poles $x_i$ and matrix residues at the poles 
$\rho_i^{\alpha \beta}=a_i^{\alpha}b_i^{\beta}$ of such solutions 
as functions of the time $t_2$ move as particles of spin generalization 
of the elliptic Calogero-Moser model (elliptic Gibbons-Hermsen model).
In this paper we establish the correspondence
with the spin elliptic Calogero-Moser model for the whole matrix KP hierarchy.
Namely, we show that the dynamics of poles and matrix residues 
of the solutions with respect to the 
$k$-th hierarchical time of the matrix KP hierarchy is  
Hamiltonian with the 
Hamiltonian $H_k$ obtained via an expansion of the spectral curve near the marked points.
The Hamiltonians are identified with the Hamiltonians of the elliptic
spin Calogero-Moser system with coordinates $x_i$ and spin degrees of freedom
$a_i^{\alpha}, \, b_i^{\beta}$. 
\end{abstract}

\end{titlepage}

\tableofcontents

\vspace{5mm}

\section{Introduction}

The Kadomtsev-Petviashvili (KP) hierarchy is an archetypal infinite hierarchy 
of compatible nonlinear differential equations with
infinitely many independent (time) variables 
${\bf t}=\{t_1, t_2, t_3, \ldots \}$.
In the Lax-Sato formalism, the main object 
is the Lax operator which is a pseudo-differential operator of the form
\beq\label{int0}
{\cal L}=\p_x +u_1\p_x^{-1}+u_2 \p_x^{-2}+\ldots
\eeq
The coefficient functions $u_i$ are dependent variables. 
Among all solutions to equations of the hierarchy,
of special interest are solutions which have a finite number of poles 
in the variable $x$ in a fundamental domain of 
the complex plane. Most general solutions of this type 
are those for which the coefficient functions $u_i$ 
are elliptic (double-periodic in the complex plane) 
functions of $x$ with poles depending on the times ${\bf t}$. 

The study of singular solutions to nonlinear integrable
equations and dynamics of their poles was initiated in the pioneering papers
\cite{AMM77,Krichever78,Krichever80,CC77}. Now it 
is a well known subject in the theory of integrable systems. 
The remarkable result is
that the poles of solutions to the KP equation 
as functions of the time $t_2$ move 
as particles of the integrable Calogero-Moser many-body system
\cite{Calogero71,Calogero75,Moser75,OP81} which is known to be integrable, i.e.
having a large number of integrals of motion in involution. Elliptic, 
trigonometric and rational solutions
correspond respectively to the Calogero-Moser systems with 
elliptic, trigonometric and rational potentials.

In the work \cite{Shiota94}, Shiota has shown that in the case of rational solutions
the correspondence between the KP equation and the rational Calogero-Moser
system can be extended to the whole KP hierarchy. 
Namely, the evolution of poles with respect to the higher time $t_m$ was considered
and it was shown that it is described by 
the higher Hamiltonian flow  
of the rational Calogero-Moser system with the Hamiltonian 
$H_m=\mbox{tr}\, L^m$, where $L$ is the Lax matrix depending on the coordinates and momenta
in a special way. 
Recently this remarkable correspondence was generalized to trigonometric 
and elliptic solutions to the KP hierarchy (see respectively 
\cite{Haine07,Z19a} and \cite{PZ21}). However, in the elliptic case this correspondence
is no longer formulated in terms of traces of the Lax matrix (which in this case depends 
on a spectral parameter). Instead, the Hamiltonian $H_m$ which governs the dynamics of poles
with respect to $t_m$ is shown to be obtained by expansion of the Calogero-Moser 
spectral curve near a distinguished marked point. It was also shown in \cite{PZ21}
that for trigonometric
and rational degenerations of elliptic solutions this construction gives the results
which agree with the previously
obtained ones for trigonometric and rational solutions. 

There exists a matrix generalization of the KP hierarchy (matrix KP hierarchy).
In matrix KP hierarchy, the coefficient functions $u_i$ in the Lax operator (\ref{int0})
are $n\! \times \! n$ matrices. 
Like the KP hierarchy, it
is an infinite set of compatible nonlinear differential equations with
infinitely many independent variables 
${\bf t}$ and matrix dependent variables. 
It is a subhierarchy of a more general multi-component ($n$-component) KP hierarchy
\cite{DJKM81,KL93,TT07,Teo11}, which has an extended set of independent variables
$
\{{\bf t}_1, {\bf t}_2, \ldots , {\bf t}_n\}$, 
${\bf t}_{\alpha}=\{t_{\alpha , 1}, t_{\alpha , 2}, t_{\alpha , 3}, \ldots \, \}$,
$\alpha = 1, \ldots , n.
$
The matrix KP hierarchy is obtained by the restriction $t_{\alpha , m}=t_m$ for all
$\alpha , m$. 

The elliptic, trigonometric and rational solutions to the matrix KP
equation were investigated in \cite{KBBT95}. In the matrix case, the data
of singular solutions include not only the positions of poles $x_i$ but also
some ``internal degrees of freedom'' which are matrix residues at the poles
(they were fixed in the scalar case). 
In the work \cite{KBBT95} it was shown that the dynamics of the data of such solutions
with respect to the time $t_2$ is isomorphic to the dynamics of a 
spin generalization of the Calogero-Moser 
system which is also known as the Gibbons-Hermsen model \cite{GH84}. It is a system of $N$
particles with coordinates $x_i$ and with internal degrees of freedom 
represented by $n$-dimensional column vectors ${\bf a}_i, {\bf b}_i$ with components
$a_i^{\alpha}$, $b_i^{\alpha}$, $\alpha =1, \ldots , n$.
The rank 1 matrices 
$\rho_i=
{\bf a}_i{\bf b}_i^T$, where ${\bf b}_i^T$ is the row vector obtained from the vector 
${\bf b}_i$ by transposition, represent matrix residues at the poles $x_i$. 
The particles pairwise interact with each other. 
The Hamiltonian of the elliptic model is
\beq\label{int2}
H=\sum_{i=1}^{{\cal N}}p_i^2-\sum_{i\neq k}
({\bf b}_i^T{\bf a}_k)({\bf b}_k^T{\bf a}_i)\wp (x_i-x_k),
\eeq
where $\wp (x)$ is the Weierstrass $\wp$-function which is the elliptic function with the 
only second order pole at $x=0$ in the fundamental domain. 
The non-vanishing Poisson brackets are
$
\{x_i, p_k\}=\delta_{ik}, \, \{a_i^{\alpha}, b_k^{\beta}\}=
\delta_{\alpha \beta}\delta_{ik}
$.
The model is known to be integrable and possessing the Lax representation with the Lax matrix
$L(\lambda )$ depending on a spectral parameter $\lambda$ lying on an elliptic curve.  

The extension of the isomorphism between rational and trigonometric solutions of the matrix
KP equation and the Gibbons-Hermsen system to the whole hierarchy was recently made in
\cite{PZ18} for rational solutions and in \cite{PZspintrig} for trigonometric ones. 
In this paper we generalize these results to elliptic solutions of the matrix
KP hierarchy. 

Our main result is that 
the dynamics of poles $x_i$ and vectors ${\bf a}_i$, ${\bf b}_i$
which parametrize matrix residues at the poles
with respect to all higher times $t_m$ of the matrix KP hierarchy
is Hamiltonian, with the corresponding Hamiltonians being higher Hamiltonians
of the spin elliptic Calogero-Moser model. We find them in terms of expansion of the 
spectral curve 
\beq\label{int3}
\det_{N\times N} \Bigl ((z+\zeta (\lambda ))I-L(\lambda )\Bigr )=0
\eeq
($\zeta (\lambda )$ is the Weierstrass $\zeta$-function)
near some distinguished marked points at infinity. The spectral curve is a covering 
of the elliptic curve, where the spectral parameter $\lambda$ lives. We show that above
the point $\lambda =0$, there are $n$ points at infinity $P_{\alpha}=(\infty_{\alpha}, 0)$,
where $z=\infty$, so that there are $n$ distinguished sheets of the covering (neighborhoods
of the points $P_{\alpha}$). In a neighborhood of the point $\lambda =0$ $n$ different 
branches of the function $z(\lambda )$ 
such that $z(\lambda )\to \infty$ as $\lambda \to 0$ are defined 
by the equation of the spectral curve. Let us denote them by $z_{\alpha}(\lambda )$ and let
$\lambda_{\alpha}(z)$ be inverse functions. Our main result is that the sum
over all branches 
$\displaystyle{\sum_{\alpha =1}^n \lambda_{\alpha}(z)}$ is the generating function 
for the Hamiltonians $H_m$:
\beq\label{int4}
\sum_{\alpha =1}^n \lambda_{\alpha}(z)=-Nz^{-1}-\sum_{m\geq 1}z^{-m-1}H_m
\eeq
or $H_m=-\displaystyle{\sum_{\alpha =1}^n\res_{\infty}\Bigl (z^m \lambda_{\alpha}(z)\Bigr )}$.
We also show that the degeneration of this construction to the rational and 
trigonometric cases allows one to reproduce the results of the papers \cite{PZ18}
and \cite{PZspintrig}.

The organization of the paper is as follows. In section 2 we remind the reader 
the main facts about the multi-component and matrix KP hierarchies. We recall both
Lax-Sato approach based on Lax equations and the bilinear (Hirota) approach based on 
the bilinear relation for the tau-function. In section 3 we introduce elliptic solutions 
and discuss the corresponding double-Bloch solutions for the wave function. Section 4
contains derivation of the dynamics of poles and residues with respect to the time $t_2$.
Following \cite{KBBT95}, we derive the equations of motion together with their Lax
representation. In section 5 we discuss properties of the spectral curve and define the
distinguished branches of the function $z(\lambda )$ around the point $\lambda =0$.
Sections 6 and 7 are devoted to derivation of the Hamiltonian dynamics of 
respectively poles and residues in the higher times $t_m$. In section 8 we find explicitly
the first two Hamiltonians using the expansion (\ref{int4}) and identify them with the 
Hamiltonians of the spin generalization of the Calogero-Moser system. Finally, in section 9
we consider the rational and trigonometric degenerations of our construction and show that 
the results of the previous works are reproduced by the new approach.

\section{The matrix KP hierarchy}

Here we briefly review the main facts about the multi-component and matrix
KP hierarchies following \cite{TT07,Teo11}.
We start from the more general multi-component KP hierarchy.
The independent variables are $n$ infinite sets of continuous ``times''
$$
{\bf t}=\{{\bf t}_1, {\bf t}_2, \ldots , {\bf t}_n\}, \qquad
{\bf t}_{\alpha}=\{t_{\alpha , 1}, t_{\alpha , 2}, t_{\alpha , 3}, \ldots \, \},
\qquad \alpha = 1, \ldots , n.
$$
It is convenient to introduce also the variable $x$ such that 
\beq\label{multi2}
\p_x =\sum_{\alpha =1}^n \p_{t_{\alpha , 1}}.
\eeq
The hierarchy is an infinite set of evolution equations in the times ${\bf t}$ for matrix
functions of the variable $x$. 

In the Lax-Sato formalism, the main object 
is the Lax operator which is a pseudo-differential operator of the form
\beq\label{multi3}
{\cal L}=\p_x +u_1\p_x^{-1}+u_2 \p_x^{-2}+\ldots
\eeq
where the coefficients $u_i=u_i(x, {\bf t})$ are $n\! \times \! n$ matrices.
The coefficient functions $u_k$ depend on $x$ and also on all the times:
$$
u_k(x, {\bf t})=u_k(x+t_{1,1}, x+t_{2,1}, \ldots , x+t_{n,1};
t_{1,2}, \ldots , t_{n,2}; \ldots ).
$$ 
Besides, there are
other matrix pseudo-differential operators ${\cal R}_1, \ldots , {\cal R}_n$ 
of the form
\beq\label{multi5}
{\cal R}_{\alpha}=E_{\alpha}+ u_{\alpha , 1}\p_x^{-1}+u_{\alpha , 2}\p_x^{-2}+\ldots ,
\eeq
where $E_{\alpha}$ is the $n \! \times \! n$ matrix with the $(\alpha , \alpha )$ element
equal to 1 and all other components equal to $0$, and $u_{\alpha , i}$ are also
$n \! \times \! n$ matrices. The operators ${\cal L}$, ${\cal R}_1, \ldots , {\cal R}_n$
satisfy the conditions 
\beq\label{multi6}
{\cal L}{\cal R}_{\alpha}={\cal R}_{\alpha}{\cal L}, \quad
{\cal R}_{\alpha}{\cal R}_{\beta}=\delta_{\alpha \beta}{\cal R}_{\alpha}, \quad
\sum_{\alpha =1}^n {\cal R}_{\alpha}=I,
\eeq
where $I$ is the unity matrix.
The Lax equations of the hierarchy which define evolution in the times read
\beq\label{multi8}
\p_{t_{\alpha , k}}{\cal L}=[B_{\alpha , k}, \, {\cal L}], \quad
\p_{t_{\alpha , k}}{\cal R}_{\beta}=[B_{\alpha , k}, \, {\cal R}_{\beta}],
\quad B_{\alpha , k} = \Bigl ({\cal L}^k{\cal R}_{\alpha}\Bigr )_+, 
\quad k=1,2,3, \ldots ,
\eeq
where $(\ldots )_+$ means the differential part of a pseudo-differential operator,
i.e. the sum of all terms with $\p_x^k$, where $k\geq 0$. 

Let us introduce the matrix pseudo-differential ``wave operator'' ${\cal W}$ with matrix elements
\beq\label{m113}
{\cal W}_{\alpha \beta} = \delta_{\alpha \beta}+\sum_{k\geq 1}
\xi^{(k)}_{\alpha \beta}(x,{\bf t})\p_x^{-k},
\eeq
where $\xi^{(k)}_{\alpha \beta}(x,{\bf t})$ are the
some matrix functions. The operators ${\cal L}$ and ${\cal R}_{\alpha}$ are obtained 
from the ``bare'' operators $I\p_x $ and $E_{\alpha}$ by `dressing'' 
by means of the wave operator:
\beq\label{multi12}
{\cal L}={\cal W}\p_x {\cal W}^{-1}, \qquad
{\cal R}_{\alpha}={\cal W}E_{\alpha} {\cal W}^{-1}.
\eeq
Clearly, there is an ambiguity in the definition of the dressing operator: it can be multiplied
from the right by any pseudo-differential operator with constant coefficients 
commuting with $E_{\alpha}$ for any $\alpha$.

A very important role in the theory is played by the wave function $\Psi$ and its adjoint
$\Psi^{\dag}$ (hereafter, ${}^{\dag}$ does not mean Hermitean conjugation).
The wave function 
is defined as a result of action of the wave operator 
to the exponential function:
\beq\label{m113a}
\Psi (x,{\bf t}; z)={\cal W}
\exp \Bigl (xzI+\sum_{\alpha =1}^n E_{\alpha}\xi ({\bf t}_{\alpha}, z)\Bigr ),
\eeq
where we use the standard notation
$$
\xi ({\bf t}_{\alpha}, z)=\sum_{k\geq 1}t_{\alpha , k}z^k.
$$
By definition, the operators $\p_x^{-k}$ with negative powers act 
to the exponential function as $\p_x^{-k}e^{xz}=z^{-k}e^{xz}$. 
The wave function depends on the spectral parameter $z$ which does not enter the auxiliary
linear problems explicitly. 
The adjoint wave function is introduced by the formula
\beq\label{m113b}
\Psi ^{\dag} (x,{\bf t}; z)=\exp \Bigl (-xzI-\sum_{\alpha =1}^n 
E_{\alpha}\xi ({\bf t}_{\alpha}, z)\Bigr )
{\cal W}^{-1}.
\eeq
Here we use the convention that 
the operators $\p_x$ 
which enter ${\cal W}^{-1}$ act to the left rather than to the right, 
the left action being defined as 
$f\stackrel{\leftarrow}{\p_x} \equiv -\p_x f$. Clearly, the expansion of the wave function
as $z\to \infty$ is as follows:
\beq\label{multi13}
\Psi_{\alpha \beta} (x,{\bf t}; z)=e^{xz+\xi ({\bf t}_{\beta}, z)}
\Bigl (\delta_{\alpha \beta}+\xi^{(1)}_{\alpha \beta}z^{-1}+
\xi^{(2)}_{\alpha \beta}z^{-2}+\ldots \Bigr ).
\eeq

As is proved in \cite{Teo11}, the wave function 
satisfies the linear
equations
\beq\label{m13c}
\p_{t_{\alpha , m}}\Psi (x,{\bf t}; z)=B_{\alpha ,m} \Psi (x,{\bf t}; z), 
\eeq
where $B_{\alpha ,m}$ is the differential operator (\ref{multi8}), i.e.
$
B_{\alpha ,m}= \Bigl ({\cal W} E_{\alpha}\p_x^m {\cal W}^{-1}\Bigr )_+
$
and the adjoint wave function satisfies the transposed equations
\beq\label{m13e}
-\p_{t_{\alpha , m}}\Psi^{\dag} (x,{\bf t}; z)=\Psi^{\dag} (x,{\bf t}; z) B_{\alpha ,m} .
\eeq
Again, the operator
$B_{\alpha , m}$ here acts to the left.
In particular, it follows from
(\ref{m13c}), (\ref{m13e}) at $m=1$ that
\beq\label{m13d}
\sum_{\alpha =1}^{n}\p_{t_{\alpha , 1}}\Psi (x,{\bf t}; z)=\p_x \Psi (x,{\bf t}; z),
\qquad
\sum_{\alpha =1}^{n}\p_{t_{\alpha , 1}}\Psi^{\dag} (x,{\bf t}; z)=\p_x \Psi^{\dag} (x,{\bf t}; z),
\eeq
so the vector field $\p_x$ can be naturally identified with the vector field
$\displaystyle{\sum_{\alpha }\p_{t_{\alpha , 1}}}$.

Another approach to the multi-component KP hierarchy is provided by the 
bilinear formalism. 
In the bilinear formalism, 
the dependent variables are the tau-function $\tau (x,{\bf t})$ and tau-functions
$\tau _{\alpha \beta}(x,{\bf t})$ such that $\tau _{\alpha \alpha}(x,{\bf t})=
\tau (x, {\bf t})$ for any $\alpha$. 
The $n$-component KP hierarchy is the infinite set of bilinear equations
for the tau-functions which are encoded in the basic bilinear
relation
\beq\label{m5}
\sum_{\nu =1}^n \epsilon_{\alpha \nu}\epsilon_{\beta \nu}
\oint_{C_{\infty}}\! dz \, 
z^{\delta_{\alpha \nu}+\delta_{\beta \nu}-2}
e^{\xi ({\bf t}_{\nu}-{\bf t}_{\nu}', \, z)}
\tau _{\alpha \nu} \left (x,{\bf t}-[z^{-1}]_{\nu}\right )
\tau _{\nu \beta}\left (x,{\bf t}'+[z^{-1}]_{\nu}\right )=0
\eeq 
valid for any ${\bf t}$, ${\bf t}'$. Here $\epsilon_{\alpha \beta}$ is a sign factor:
$\epsilon_{\alpha \beta}=1$ if $\alpha \leq \beta$, $\epsilon_{\alpha \beta}=-1$
if $\alpha >\beta$.
In (\ref{m5}) we use the following standard
notation:
$$
\left ({\bf t}\pm [z^{-1}]_{\gamma}\right )_{\alpha k}=t_{\alpha , k}\pm
\delta_{\alpha \gamma} \frac{z^{-k}}{k}.
$$
The integration contour $C_{\infty}$ is a big circle around $\infty$.

The tau-functions are universal dependent variables of the hierarchy.
All other objects including
the coefficient functions $u_i$ of the Lax operator and the wave functions can be 
expressed in terms of them. In particular, for the wave function and its adjoint we have:
\beq\label{m2}
\begin{array}{l}
\displaystyle{\Psi_{\alpha \beta}(x,{\bf t};z)=
\epsilon_{\alpha \beta}\,
\frac{\tau_{\alpha \beta} \left (x,
{\bf t}-[z^{-1}]_{\beta}\right )}{\tau (x,{\bf t})}\,
z^{\delta_{\alpha \beta}-1}e^{\xi ({\bf t}_{\beta}, z)},
}
\\ \\
\displaystyle{\Psi_{\alpha \beta}^{\dag}(x,{\bf t};z)=
\epsilon_{\beta \alpha}\,
\frac{\tau _{\alpha \beta}\left (x,
{\bf t}+[z^{-1}]_{\alpha}\right )}{\tau (x,{\bf t})}\,
z^{\delta_{\alpha \beta}-1}e^{-\xi ({\bf t}_{\alpha}, z)}
}
\end{array}
\eeq
Note that the bilinear relation (\ref{m5}) can be written in the form
\beq\label{m3}
\oint_{C_{\infty}}\! dz \, \Psi (x,{\bf t};z)\Psi^{\dag} (x,{\bf t}';z)=0.
\eeq

The coefficient $\xi_{\alpha \beta}^{(1)}(x,{\bf t})$ 
plays an important role in what follows. Equations (\ref{m2}) imply that this coefficient
is expressed through the tau-functions as
\beq\label{m2a}
\xi_{\alpha \beta}^{(1)}(x,{\bf t})=\left \{
\begin{array}{l}
\displaystyle{\epsilon_{\alpha \beta}\, 
\frac{\tau_{\alpha \beta}(x, {\bf t})}{\tau (x, {\bf t})}, \quad \alpha \neq \beta ,}
\\ \\
\displaystyle{
-\frac{\p_{t_{\beta}}\tau (x, {\bf t})}{\tau (x, {\bf t})}, \quad \,\, \alpha =\beta .}
\end{array}
\right.
\eeq

Let us point out a useful corollary of the bilinear
relation (\ref{m5}).
Differentiating it with respect to $t_{\kappa , m}$ and putting ${\bf t}'={\bf t}$ after this, we 
obtain:
\beq\label{m11b}
\frac{1}{2\pi i}
\oint_{C_{\infty}}dz \, z^m \Psi_{\alpha \kappa}(x,{\bf t}; z)
\Psi^{\dag}_{\kappa \beta}(x,{\bf t}; z)=-\p_{t_{\kappa , m}}\xi_{\alpha \beta}^{(1)}(x,{\bf t})
\eeq
or, equivalently, 
\beq\label{m11c}
\res_{\infty}\Bigl (z^m \Psi_{\alpha \kappa}\Psi^{\dag}_{\kappa \beta}
\Bigr ) =-\p_{t_{\kappa , m}}\xi_{\alpha \beta}^{(1)}.
\eeq
The integrand in (\ref{m11b}) should be regarded as a Laurent series in $z$ and the
residue at infinity is defined according to the convention 
$\res_{\infty}\, (z^{-k})=\delta_{k1}$.

The matrix KP hierarchy is a subhierarchy of the 
multi-component KP one.  It is obtained by
the following restriction of the independent 
variables:
$t_{\alpha , m}=t_m$ for each $\alpha$ and $m$.
The corresponding vector fields are related as
$\displaystyle{\p_{t_m}=\sum_{\alpha =1}^n \p_{t_{\alpha , m}}}$.
As is clear from (\ref{multi13}), 
the wave function for the matrix KP hierarchy has the expansion
\beq\label{m7}
\Psi_{\alpha \beta}(x,{\bf t};z)=\left (\delta_{\alpha \beta}+
\xi_{\alpha \beta}^{(1)}({\bf t})z^{-1}+O(z^{-2})\right )
e^{xz+\xi ({\bf t}, z)},
\eeq
where $\displaystyle{\xi ({\bf t}, z)=\sum_{k\geq 1}t_kz^k}$.
Equations (\ref{m13c}) imply that the
wave function of the matrix KP hierarchy and its adjoint 
satisfy the linear
equations
\beq\label{m13a}
\p_{t_m}\Psi ({\bf t}; z)=B_m \Psi ({\bf t}; z), 
\qquad
-\p_{t_m}\Psi^{\dag} ({\bf t}; z)=\Psi^{\dag} ({\bf t}; z) B_m , \qquad m\geq 1,
\eeq
where $B_m$ is the differential operator
$
B_m= \Bigl ({\cal W} \p_x^m {\cal W}^{-1}\Bigr )_+.
$
At $m=1$ we have $\p_{t_1}\Psi =\p_x \Psi$, 
so we can identify
$\displaystyle{
\p_x = \p_{t_1}=\sum_{\alpha =1}^N \p_{t_{\alpha , 1}}}
$. This means that
the evolution in the time $t_1$ is simply a shift of
the variable $x$:
$
\xi^{(k)}(x, t_1, t_2, \ldots )=\xi^{(k)}(x+t_1, t_2, \ldots ).
$
At $m=2$ equations (\ref{m13a}) turn into the linear problems
\beq\label{m14}
\p_{t_2}\Psi = \p_x^2\Psi +2V(x,{\bf t})\Psi ,  
\eeq
\beq\label{m14a}
-\p_{t_2}\Psi^{\dag} = \p_x^2\Psi^{\dag} +2\Psi^{\dag}V(x,{\bf t})
\eeq
which have the form of the matrix non-stationary Schr\"odinger equations with
\beq\label{m15}
V(x,{\bf t})=-\p_x \xi^{(1)}(x,{\bf t}).
\eeq

Summing (\ref{m11b}) over $\kappa$, we obtain an analog of (\ref{m11b}) for the
matrix KP hierarchy:
\beq\label{m11a}
\frac{1}{2\pi i}
\sum_{\nu =1}^n \oint_{C_{\infty}}dz \, z^m \Psi_{\alpha \nu}(x,{\bf t}; z)
\Psi^{\dag}_{\nu \beta}(x,{\bf t}; z)=-\p_{t_m}\xi_{\alpha \beta}^{(1)}(x,{\bf t}).
\eeq
Below we will use equations (\ref{m11b}) and (\ref{m11a}) 
for derivation of dynamics of poles and residues of elliptic solutions
in higher times.

\section{Elliptic solutions of the matrix KP hierarchy and double-Bloch functions}

Our aim is to study solutions to the matrix KP hierarchy which are elliptic functions
of the variable $x$ (and, therefore, $t_1$).
For the elliptic solutions, we take the tau-function in the form
\beq\label{t1}
\tau (x,{\bf t})= C\prod_{i=1}^{{\cal N}} \sigma (x-x_i({\bf t})),
\eeq
where
$$
\sigma (x)=\sigma (x |\, \omega , \omega ')=
x\prod_{s\neq 0}\Bigl (1-\frac{x}{s}\Bigr )\, e^{\frac{x}{s}+\frac{x^2}{2s^2}},
\quad s=2\omega m+2\omega ' m' \quad \mbox{with integer $m, m'$}
$$ 
is the Weierstrass 
$\sigma$-function with quasi-periods $2\omega$, $2\omega '$ such that 
${\rm Im} (\omega '/ \omega )>0$. 
It is connected with the Weierstrass 
$\zeta$- and $\wp$-functions by the formulas $\zeta (x)=\sigma '(x)/\sigma (x)$,
$\wp (x)=-\zeta '(x)=-\p_x^2\log \sigma (x)$.
The monodromy properties of the function $\sigma (x)$ 
are
\beq\label{e1a}
\sigma (x+2\omega )=-e^{2\zeta (\omega ) (x+\omega )}\sigma (x), \quad
\sigma (x+2\omega ' )=-e^{2\zeta (\omega ' ) (x+\omega ' )}\sigma (x),
\eeq
where the constants $\zeta (\omega )$, $\zeta (\omega ' )$ are  related by
$\zeta (\omega )\omega ' -\zeta (\omega ') \omega =\pi i /2$.
The $N$ zeros $x_i$ of (\ref{t1}) are assumed to be all distinct. 

We also assume that the tau-functions $\tau_{\alpha \beta}$ at $\alpha \neq \beta$
have the form
\beq\label{t1a}
\tau_{\alpha \beta} (x,{\bf t})= C_{\alpha \beta}\prod_{i=1}^{{\cal N}} 
\sigma (x-x_i^{(\alpha \beta )}({\bf t}))
\eeq
with
\beq\label{t1b}
\sum_i x_i({\bf t}) = \sum_i x_i^{(\alpha \beta )}({\bf t}) 
\quad \mbox{for all $\alpha , \beta$}.
\eeq
The consistency of this assumption is justified below. 

Equation (\ref{m2a}) together with the condition (\ref{t1b}) implies that
$V(x, {\bf t})=-\p_x \xi^{(1)}$ in the linear problem (\ref{m14}) is an
elliptic function of $x$. Therefore, one can find solutions to (\ref{m14})
which are {\it double-Bloch functions}.
The double-Bloch function satisfies the monodromy properties
$\Psi_{\alpha \beta} (x+2\omega )=B_{\beta}\Psi_{\alpha \beta} (x)$, 
$\Psi_{\alpha \beta} (x+2\omega ' )=B_{\beta}'\Psi_{\alpha \beta} (x)$ with some
Bloch multipliers $B_{\beta}$, $B_{\beta}'$. 
The Bloch multipliers of the wave function
(\ref{m2}) are:
\beq\label{e3}
\begin{array}{l}
\displaystyle{
B_{\beta}=\exp \Bigl (2\omega z-2\zeta (\omega )\sum_i (e^{-D_{\beta}(z)}\! -\! 1)x_i )\Bigr ),}
\\ \\
\displaystyle{
B_{\beta}'=\exp \Bigl (2\omega ' z -2\zeta (\omega ' )\sum_i (e^{-D_{\beta}(z)}\! -\! 1)x_i)\Bigr ),}
\end{array}
\eeq
where the differential operator $D_{\beta}(z)$ is
\beq\label{D(z)}
D_{\beta}(z)=\sum_{k\geq 1}\frac{z^{-k}}{k}\, \p_{t_{\beta ,k}}.
\eeq
Since the right hand side of (\ref{m11a}) is an elliptic function of $x$, 
the Bloch multipliers of the adjoint wave function should be $1/B_{\alpha}$, 
$1/B_{\alpha}'$: $\Psi^{\dag}_{\alpha \beta} (x+2\omega )=(B_{\alpha})^{-1}
\Psi^{\dag}_{\alpha \beta} (x)$, $\Psi^{\dag}_{\alpha \beta} (x+2\omega ' )=(B'_{\alpha})^{-1}
\Psi^{\dag}_{\alpha \beta} (x)$.

Any non-trivial double-Bloch function (i.e. the one which is not just an exponential
function) must have at least one pole in $x$ in the fundamental domain. 
Let us introduce the 
elementary double-Bloch function $\Phi (x, \lambda )$ having just one pole 
in the fundamental domain and defined as
\beq\label{Phi}
\Phi (x, \lambda )=\frac{\sigma (x+\lambda )}{\sigma (\lambda )\sigma (x)}\,
e^{-\zeta (\lambda )x}
\eeq
(here $\zeta (\lambda )$ is the Weierstrass $\zeta$-function).
The monodromy properties of the function $\Phi$ follow from (\ref{e1a}):
$$
\Phi (x+2\omega , \lambda )=e^{2(\zeta (\omega )\lambda - \zeta (\lambda )\omega )}
\Phi (x, \lambda ),
$$
$$
\Phi (x+2\omega ' , \lambda )=e^{2(\zeta (\omega ' )\lambda - \zeta (\lambda )\omega ' )}
\Phi (x, \lambda ).
$$
We see that it is indeed a double-Bloch function.
It has a single simple pole in the fundamental domain
at $x=0$ with residue 1: 
$$
\Phi (x, \lambda )=\frac{1}{x} -\frac{1}{2}\, \wp (\lambda )x  +\ldots , \qquad 
x\to 0,
$$
It is easy to see that $\Phi (x, \lambda )$ is an elliptic function of $\lambda$. 
The expansion as $\lambda \to 0$ is
\beq\label{Phi1}
\begin{array}{c}
\Phi (x, \lambda )=\left (\lambda^{-1}+ \zeta (x)+\frac{1}{2} \, (\zeta^2(x)-\wp (x))\lambda +
O(\lambda^2)\right )e^{-x/\lambda}.
\end{array}
\eeq
We will also need the $x$-derivatives 
$\Phi '(x, \lambda )=\p_x \Phi (x, \lambda )$, 
$\Phi ''(x, \lambda )=\p^2_x \Phi (x, \lambda )$.

It is clear from (\ref{m2}) that the wave functions $\Psi$, $\Psi^\dag$
(and thus the coefficient $\xi^{(1)}$), 
as functions of $x$, 
have simple poles at $x=x_i$. 
It is shown in \cite{PZ18} that the residues 
at these poles are matrices of rank $1$. We parametrize the residues
of $\xi^{(1)}$ through the 
column vectors
${\bf a}_i =(a_i^1, a_i^2, \ldots , a_i^n)^T$,
${\bf b}_i =(b_i^1, b_i^2, \ldots , b_i^n)^T$ ($T$ means transposition):
\beq\label{t3a}
\xi_{\alpha \beta}^{(1)}=S_{\alpha \beta}-\sum_i a_i^{\alpha}b_i^{\beta}\zeta (x-x_i),
\eeq
where $S_{\alpha \beta}$ does not depend on $x$.
Therefore,
\beq\label{t3b}
V(x, {\bf t})=-\sum_i a_i^{\alpha}b_i^{\beta}\wp (x-x_i).
\eeq
The components of the vectors ${\bf a}_i$, ${\bf b}_i$ are going to be
spin variables of the elliptic Gibbons-Hermsen model. 

One can expand the wave functions
using the elementary double-Bloch functions as follows: 
\beq\label{t3}
\Psi_{\alpha \beta}=e^{k_{\beta} x+\xi ({\bf t}, z)}\sum_i a_i^{\alpha}c_i^{\beta}
\Phi (x-x_i, \lambda_{\beta}),
\eeq
\beq\label{t4}
\Psi^{\dag}_{\alpha \beta}=e^{-k_{\alpha} x-\xi ({\bf t}, z)}\sum_i
c_i^{*\alpha}b_i^{\beta}\Phi (x-x_i, -\lambda_{\alpha}),
\eeq
where $c_i^{\alpha}$, $c_i^{*\alpha}$ are components of some $x$-independent vectors
${\bf c}_i=(c_i^1,  \ldots , c_i^n)^T$, 
${\bf c}^{*}_i=(c^{*1}_i,  \ldots , c^{*n}_i)^T$.
This is similar to expansion of a rational function
in a linear combination of simple fractions.

One can see that (\ref{t3}) is 
a double-Bloch function with Bloch multipliers
\beq\label{e6}
B_{\beta}=e^{2\omega (k_{\beta}-\zeta (\lambda _{\beta})) + 
2\zeta (\omega )\lambda_{\beta} }, \qquad
B_{\beta} '=e^{2\omega ' (k_{\beta}-\zeta (\lambda_{\beta} )) + 
2\zeta (\omega ' )\lambda_{\beta} }
\eeq
and (\ref{t4}) has Bloch multipliers $(B_{\alpha})^{-1}$ and $(B_{\alpha}')^{-1}$. 
These Bloch multipliers
should coincide with (\ref{e3}). 
Therefore, comparing (\ref{e3}) with (\ref{e6}), we get
$$
2\omega \Bigl (k_{\beta}-z-\zeta (\lambda_{\beta} )\Bigr )+
2\zeta (\omega )\Bigl (\lambda_{\beta} +(e^{-D_{\beta}(z)}-1)\sum_i x_i\Bigr )=
2\pi i n,
$$
$$
2\omega ' \Bigl (k_{\beta}-z-\zeta (\lambda_{\beta} )\Bigr )+
2\zeta (\omega ' )\Bigl (\lambda_{\beta} +(e^{-D_{\beta}(z)}-1)\sum_i x_i\Bigr )=
2\pi i n'
$$
with some integer $n, n'$. These equations can be regarded as a linear system.
The solution is
$$
k_{\beta}-z-\zeta (\lambda_{\beta})=2n'\zeta (\omega)-2n \zeta (\omega '),
$$
$$
\lambda_{\beta} +(e^{-D_{\beta}(z)}-1)\sum_i x_i=2n\omega ' -2n'\omega .
$$
Shifting $\lambda_{\beta}$ by a suitable vector of the lattice spanned by $2\omega$, $2\omega '$,
we can 
represent the connection between the spectral parameters $k_{\beta},z, \lambda_{\beta}$ 
in the form
\beq\label{e7}
\left \{\begin{array}{l}
k_{\beta}=z+\zeta (\lambda_{\beta} ),
\\ \\
\displaystyle{\lambda_{\beta} = (1-e^{-D_{\beta}(z)}) \sum_i x_i}.
\end{array} \right.
\eeq
These two equations for three spectral parameters $k_{\beta}, z, \lambda_{\beta}$ 
determine the spectral curve, with the index $\beta$ numbering different sheets of it. 
Another description of the same spectral curve is obtained below as 
the spectral curve of the spin generalization of the Calogero-Moser system 
(it is given by the characteristic polynomial of the Lax matrix
$L(\lambda )$ for the spin Calogero-Moser system). 
As we shall see below, it has the form $R(k,\lambda)=0$, where
$R(k,\lambda)$ is a polynomial in $k$ whose coefficients are elliptic functions of
$\lambda$. 
These coefficients are integrals of motion in involution. The spectral curve
in the form $R(k,\lambda)=0$ appears if one excludes $z$ from the equations (\ref{e7}).
Equivalently, one can represent the spectral curve as a relation connecting 
$z$ and $\lambda_{\beta}$:
\beq\label{e7a}
R(z+\zeta (\lambda_{\beta} ), \lambda_{\beta} )=0.
\eeq
The function $z(\lambda )$ defined by this equation is multivalued, $z_{\beta}(\lambda )$
being different branches of this function. Then the function $\lambda_{\beta}(z)$ is the inverse
function to the $z_{\beta}(\lambda )$.
Using the same arguments as in \cite{PZ21}, one can see that
the second equation in (\ref{e7}) can be written as
\beq\label{e8}
\lambda_{\beta}(z) = D_{\beta}(z)\sum_i x_i=\sum_{j\geq 1}\frac{z^{-j}}{j}\, V^{(\beta )}_j,
\quad V^{(\beta )}_j=\p_{t_{\beta , j}}\sum_i x_i,
\eeq
where $\lambda_{\beta}(z)$ should be understood as expansion of the 
$\beta$-th branch of the function $\lambda (z)$ in negative powers of $z$ near $z=\infty$. 

\section{Dynamics of poles and residues in $t_2$}

We first consider the dynamics of the poles and residues
with respect to the time $t_2$. Following Krichever's approach, 
we consider the linear problems
(\ref{m14}), (\ref{m14a}),
and substitute the pole ansatz (\ref{t3}), (\ref{t4}) for the wave functions. 

Consider first the equation for $\Psi$. 
After the substitution, we see that the expression has poles at $x=x_i$
up to the third order. 
Equating coefficients at the poles of different orders at $x=x_i$,
we get the conditions:
\begin{itemize}
\item
At $\frac{1}{(x-x_i)^3}$: \phantom{a} $b_i^{\nu}a_i^{\nu}=1$;
\item
At $\frac{1}{(x-x_i)^2}$: \phantom{a} $\displaystyle{
-\frac{1}{2}\, \dot x_i c_i^{\beta}-\sum_{j\neq i}
b_i^{\nu}a_j^{\nu}c_j^{\beta}\Phi (x_i\! -\! x_j, \lambda_{\beta})=
k_{\beta}c_i^{\beta}}$;
\item
At $\frac{1}{x-x_i}$: $\phantom{aaaa} \p_{t_2}(
a_i^{\alpha}c_i^{\beta})=(k_{\beta}^2-z^2 
+\wp (\lambda_{\beta} )) a_i^{\alpha}c_i^{\beta}$
$$
\phantom{aaaaaaaaaaaaa}
-2\sum_{j\neq i}a_i^{\alpha}b_i^{\nu}a_{j}^{\nu}c_j^{\beta}\Phi '(x_i-x_j, \lambda_\beta)
-2c_i^{\beta}\sum_{j\neq i}a_i^{\nu}b_j^{\nu}a_j^{\alpha}\wp (x_i-x_j),
$$
\end{itemize}
where dot means $t_2$-derivative. Here and below summation over repeated Greek indices
numbering components of vectors from 1 to $n$ is implied, unless otherwise stated.
Similar calculations for the linear problem for $\Psi^{\dag}$ lead to the conditions
\begin{itemize}
\item
At $\frac{1}{(x-x_i)^3}$: \phantom{a} $b_i^{\nu}a_i^{\nu}=1$ (the same as above);
\item
At $\frac{1}{(x-x_i)^2}$: \phantom{a} $\displaystyle{
-\frac{1}{2}\, \dot x_i c_i^{*\alpha}- \sum_{j\neq i}
c_j^{*\alpha}b_j^{\nu}a_i^{\nu}\Phi (x_j\! -\! x_i, \lambda_{\alpha})
=k_{\alpha}c_i^{*\alpha}}$;
\item
At $\frac{1}{x-x_i}$:
$\phantom{aaaa} \p_{t_2}(
c_i^{*\alpha}b_i^{\beta})=-(k_{\beta}^2-z^2 
+\wp (\lambda_{\alpha} )) c_i^{*\alpha}b_i^{\beta}$
$$
\phantom{aaaaaaaaaaaaa}
+2\sum_{j\neq i}c_j^{*\alpha}b_j^{\nu}a_{i}^{\nu}b_i^{\beta}\Phi '(x_j-x_i, \lambda_\alpha )
+2c_i^{*\alpha}\sum_{j\neq i}b_i^{\nu}a_j^{\nu}b_j^{\beta}\wp (x_i-x_j).
$$
\end{itemize}
Here we have used the obvious property $\Phi (x, -\lambda )=-\Phi (-x, \lambda )$.

The conditions coming from the
third order poles are constraints on the vectors ${\bf a}_i$, ${\bf b}_i$. 
The other conditions can be written in the matrix form
\beq\label{t6}
\left \{
\begin{array}{l}
(k_{\beta}I-L(\lambda_{\beta})){\sf c}^{\beta}=0
\\ \\
\dot {\sf c}^{\beta}=M(\lambda_{\beta}){\sf c}^{\beta},
\end{array}
\right.
\eeq
\beq\label{t7}
\left \{
\begin{array}{l}
{\sf c}^{*\alpha}(k_{\alpha}I-L(\lambda_{\alpha}))=0
\\ \\
\dot {\sf c}^{*\alpha}={\sf c}^{*\alpha}M^{*}(\lambda_{\alpha})
\end{array}
\right.
\eeq
(no summation over $\alpha , \beta$),
where ${\sf c}^{\beta}=(c^{\beta}_1, \ldots c^{\beta}_{N})^T$,
${\sf c}^{*\alpha}=(c^{*\alpha}_1, \ldots c^{*\alpha}_N)$,
are $N$-dimensional vectors, $I$ is the unity matrix, and
$L(\lambda )$, $M(\lambda )$, $M^{*}(\lambda )$
are $N\! \times \! N$ matrices of the form
\beq\label{t8}
L_{ij}(\lambda )=-\frac{1}{2}\, \dot x_i \delta_{ij}-(1-\delta_{ij})
b_i^{\nu}a_j^{\nu}\Phi (x_i-x_j, \lambda ),
\eeq
\beq\label{t9}
M_{ij}(\lambda )=(k^2-z^2+\wp (\lambda ) -\Lambda_i)\delta_{ij}-2(1-\delta_{ij})
b_i^{\nu}a_j^{\nu}\Phi '(x_i-x_j, \lambda ),
\eeq
\beq\label{t10}
M^{*}_{ij}(\lambda )=-(k^2-z^2+\wp (\lambda ) -\Lambda^{*}_i)\delta_{ij}+2(1-\delta_{ij})
b_i^{\nu}a_j^{\nu}\Phi '(x_i-x_j, \lambda ).
\eeq
Here
\beq\label{t11}
\Lambda_i=\frac{\dot a_i^{\alpha}}{a_i^{\alpha}}+2
\sum_{j\neq i}\frac{a_j^{\alpha}b_j^{\nu}a_i^{\nu}}{a_i^{\alpha}}\, \wp (x_i-x_j),
\quad
-\Lambda_i^*=\frac{\dot b_i^{\alpha}}{b_i^{\alpha}}-2
\sum_{j\neq i}\frac{ b_i^{\nu}a_j^{\nu}b_j^{\alpha}}{b_i^{\alpha}}\, \wp (x_i-x_j)
\eeq
do not depend on the index $\alpha$ (there is summation over $\nu$
but no summation over $\alpha$). In fact one can see that $\Lambda_i=\Lambda_i^*$,
so that $M^{*}(\lambda )=-M(\lambda )$.
Indeed, multiplying equations (\ref{t11}) by $a_i^{\alpha}b_i^{\alpha}$
(no summation here!), summing over 
$\alpha$ and summing the two equations, we get $\Lambda_i-\Lambda_i^*=\p_{t_2}(
a_i^{\alpha}b_i^{\alpha})=0$ by virtue of the constraint
$a_i^{\alpha}b_i^{\alpha}=1$. 

Differentiating the first equation in (\ref{t6}) by $t_2$, we get
the compatibility condition of equations (\ref{t6}):
\beq\label{t12}
(\dot L+[L,M]){\sf c}^{\beta}=0.
\eeq
One can see, taking into account equations (\ref{t11}), which we write here in the form
\beq\label{t13}
\dot a_i^{\alpha}=\Lambda_i a_i^{\alpha}-2
\sum_{j\neq i}a_j^{\alpha}b_j^{\nu}a_i^{\nu}\wp (x_i-x_j),
\eeq
\beq\label{t13a}
\dot b_i^{\alpha}=-\Lambda_i b_i^{\alpha}+2
\sum_{j\neq i}b_i^{\nu}a_j^{\nu}b_j^{\alpha}\wp (x_i-x_j)
\eeq
(in this form they are equations of motion for the spin degrees of freedom)
that the off-diagonal elements of the matrix $\dot L+[L,M]$ are equal to zero.
Vanishing of the diagonal elements yields equations of motion for the poles $x_i$:
\beq\label{t14}
\ddot x_i=4\sum_{j\neq i}
b_i^{\mu}a_k^{\mu}
b_k^{\nu}a_i^{\nu}\wp '(x_i-x_j).
\eeq
The gauge transformation $a_i^{\alpha}\to a_i^{\alpha}q_i$,
$b_i^{\alpha}\to b_i^{\alpha}q_i^{-1}$ with
$\displaystyle{q_i=\exp \Bigl (\int^{t_2}\Lambda_i dt\Bigr )}$ eliminates 
the terms with $\Lambda_i$ in (\ref{t13}), (\ref{t13a}), so we can put
$\Lambda_i=0$. This gives the equations of motion
\beq\label{t15}
\dot a_i^{\alpha}=-2
\sum_{j\neq i}a_j^{\alpha}b_j^{\nu}a_i^{\nu}\wp (x_i-x_j),
\quad
\dot b_i^{\alpha}=2
\sum_{j\neq i}b_i^{\nu}a_j^{\nu}b_j^{\alpha}\wp (x_i-x_j).
\eeq
Together with (\ref{t14}) they are equations of motion of the elliptic 
Gibbons-Hermsen model.
Their Lax representation is given by the matrix equation $\dot L=[M,L]$. It states that
the time evolution of the Lax matrix is an isospectral transformation. It follows 
that the quantities $\mbox{tr}\, L^m(\lambda )$ are integrals of motion. In particular,
\beq\label{t16}
H_2=\sum_{i=1}^{N}p_i^2-\sum_{i\neq j}
b_i^{\mu}a_j^{\mu} b_j^{\nu}a_i^{\nu}\wp (x_i-x_j)=
\mbox{tr}\, L^2(\lambda ) +\mbox{const}
\eeq
is the Hamiltonian of the elliptic Gibbons-Hermsen model. Equations of motion (\ref{t14}),
(\ref{t15}) are equivalent to the Hamiltonian equations
\beq\label{t17}
\dot x_i=\frac{\p H_2}{\p p_i}, \quad \dot p_i=-\frac{\p H_2}{\p x_i},
\quad
\dot a_i^{\alpha}=\frac{\p H_2}{\p b_i^{\alpha}}, \quad
\dot b_i^{\alpha}=-\frac{\p H_2}{\p a_i^{\alpha}}.
\eeq
We see that $\dot x_i =2p_i$ and the Lax matrix is expressed through the momenta as follows:
\beq\label{t18}
L_{ij}(\lambda )=-p_i \delta_{ij}-(1-\delta_{ij})b_i^{\nu}a_j^{\nu}\Phi (x_i-x_j, \lambda ).
\eeq

As we shall see, the higher time flows are also Hamiltonian with the Hamiltonians
being linear combinations of spectral invariants of the Lax matrix, i.e. linear combinations
of traces of its powers $\mbox{tr}\, L^j(\lambda )$. It is not difficult to see that
\beq\label{t19}
G^{\alpha \beta}=\sum_i a_i^{\alpha}b_i^{\beta}
\eeq
are integrals of motion for all time flows: $\p_{t_m}G^{\alpha \beta}=0$. Indeed, we have
$$
\p_{t_m}\Bigl (\sum_i a_i^{\alpha}b_i^{\beta}\Bigr )=
\sum_i \left ( b_i^{\beta}\frac{\p H_m}{\p b_i^{\alpha}}-
a_i^{\alpha}\frac{\p H_m}{\p a_i^{\beta}}\right )
$$
and this is zero because $H_m$ is a linear combination of 
$\mbox{tr}\, L^j(\lambda )$ and
$$
\sum_i \left (b_i^{\beta}\mbox{tr}\, \Bigl (\frac{\p L}{\p b_i^{\alpha}}\, L^{j-1}\Bigr )
-a_i^{\alpha}\, \mbox{tr}\, \Bigl (\frac{\p L}{\p a_i^{\beta}}\, L^{j-1}\Bigr )\right )
$$
$$
=\sum_i \sum_{l,k}\left (b_i^{\beta}\, \frac{\p L_{lk}}{\p b_i^{\alpha}}\, L_{kl}^{j-1}-
a_i^{\alpha}\, \frac{\p L_{lk}}{\p a_i^{\beta}}\, L_{kl}^{j-1}\right )
$$
$$
=\sum_i \sum_{l\neq k}(\delta_{ik}-\delta_{il})b_l^{\beta}a_k^{\alpha}\Phi (x_l-x_k)
L_{kl}^{j-1}=0.
$$

A simple lemma from linear algebra states that eigenvalues $\nu_{\alpha}$ of the 
$n\times n$ matrix $G$ (\ref{t19}) coincide with nonzero eigenvalues of the rank $n$
$N\times N$ matrix $F$ with matrix elements
\beq\label{t20}
F_{ij}=b_i^{\nu}a_j^{\nu}
\eeq
(we assume that $n\leq N$). Indeed, consider the rectangular $N\times n$ matrix 
${\bf A}_{i\alpha}=
a_i^{\alpha}$ and ${\bf B}_{i \beta}=b_i^{\beta}$, then 
$G={\bf A}^T {\bf B}$, $F={\bf B}{\bf A}^T$ and a straightforward verification shows that
traces of all powers of these matrices coincide: $\mbox{tr}\, G^m=\mbox{tr}\, F^m$ for all
$m\geq 1$. This means that their nonzero eigenvalues also coincide. Note that
$\displaystyle{\mbox{tr}\, G=\sum_{\alpha =1}^n \nu_{\alpha}=N}$. 

\section{The spectral curve}

The first of the equations (\ref{t6}) determines a connection between
the spectral parameters $k=k_{\beta}, \lambda=\lambda_{\beta}$ which 
is the equation of the spectral curve:
\beq\label{spec1}
R(k, \lambda ):=\det \Bigl (kI-L(\lambda )\Bigr )=0.
\eeq
As we already mentioned, the spectral curve 
is an integral of motion.
The matrix
$L=L(\lambda )$ has an essential singularity at $\lambda =0$. It can be 
represented in the form $L=V\tilde L V^{-1}$, where 
$V$ is the diagonal matrix $V_{ij}=\delta_{ij}
e^{-\zeta (\lambda )x_i}$.
Matrix elements of 
$\tilde L$ do not have any
essential singularity in $\lambda$. We conclude that
$$
R(k, \lambda )=\sum_{m=0}^{N}R_m(\lambda )k^m,
$$
where the coefficients $R_m(\lambda )$ are elliptic functions of $\lambda$ with poles
at $\lambda =0$.
They can be represented as linear combinations of the $\wp$-function and
its derivatives, coefficients
of this expansion being integrals of motion. Fixing their values, we obtain
an algebraic curve $\Gamma$ which is an 
$N$-sheet covering of the initial elliptic curve ${\cal E}$ realized as a factor
of the complex plane with respect to the lattice generated by $2\omega$, $2\omega '$.

In a neighborhood of the point $\lambda =0$   
the matrix $\tilde L(\lambda)$
can be represented as
$$
\tilde L(\lambda )=\lambda ^{-1}(I-F)+O(1),
$$ 
where
$F$ is the rank $n$ matrix (\ref{t20}) (recall that 
$n\leq N$). This matrix has $N-n$ vanishing eigenvalues and $n$ nonzero eigenvalues 
$\nu_{\alpha}$, $\alpha =1, \ldots , n$. They are time-independent quantities because
as we have shown above they coincide with eigenvalues of the matrix $G$ (\ref{t19}) which is
an integral of motion. 
Therefore, we can write
$$
\det \Bigl (kI-L(\lambda )\Bigr )=\prod_{\alpha =1}^n \Bigl (k-(1-\nu_{\alpha})\lambda^{-1}-
h_{\alpha}(\lambda )\Bigr )\prod_{j=n+1}^N \Bigl (k-\lambda^{-1}-
h_{j}(\lambda )\Bigr ),
$$
where $h_{\alpha}$, $h_{j}$ are regular functions of $\lambda$ near $\lambda =0$. 
This means that the function $k$
has simple poles on all sheets at the points of the curve
$\Gamma$ located 
above $\lambda =0$. Now,
recalling the connection between $k$ and $z$ given by the first equation in (\ref{e7}), we
have
\beq\label{spec2}
\det \Bigl ((z+\zeta (\lambda ))I-L(\lambda )\Bigr )=\prod_{\alpha =1}^n \Bigl (z+
\nu_{\alpha}\lambda^{-1}-
h_{\alpha}(\lambda )\Bigr )\prod_{j=n+1}^N \Bigl (z-
h_{j}(\lambda )\Bigr ).
\eeq
We see that $n$ sheets of the curve $\Gamma$ lying above a neighborhood of the point 
$\lambda =0$ are distinguished. There are $n$ points at infinity above $\lambda =0$:
$P^{(\infty )}_1=(\infty _1, 0), \ldots P^{(\infty )}_n=(\infty _n, 0)$. 
In the vicinity of the point $P^{(\infty )}_{\alpha}$
the function $\lambda =\lambda_{\alpha}(z)$ has the following expansion:
\beq\label{spec3}
\lambda =\lambda_{\alpha}(z)=-\nu_{\alpha}z^{-1} +O(z^{-2}).
\eeq
As it is shown in \cite{KBBT95}, the points $P^{(\infty )}_{\alpha}\in \Gamma$ are the 
marked points, where the Baker-Akhiezer function on the spectral curve 
has essential singularities. 

With the expansion (\ref{spec3}) at hand, we can make a more detailed identification
of the wave function (\ref{m2}) with the expansion (\ref{m7}) 
and the wave function (\ref{t3}). The expansion of the function (\ref{t3}) 
as $\lambda_{\beta}\to 0$ yields
$$
\Psi_{\alpha \beta}=e^{zx+\xi ({\bf t}, z)}\sum_i
\Bigl (a_i^{\alpha}b_i^{\beta}\nu_{\beta}^{-1} +\lambda_{\beta}\nu_{\beta}^{-1}
(a_i^{\alpha}d_{i}^{\beta} +a_i^{\alpha}b_i^{\beta}\zeta (x-x_i)) +O(\lambda_{\beta}^2)\Bigr ),
$$
where we took into account that the identification implies the expansion
\beq\label{spec4}
c_i^{\beta}=\nu_{\beta}^{-1}\lambda_{\beta}^{-1}e^{-x_i\zeta (\lambda )}
\Bigl (b_i^{\beta}+\lambda_{\beta} d_i^{\beta}+O(\lambda_{\beta}^2)\Bigr ), 
\quad \lambda_{\beta}\to 0.
\eeq
Therefore, taking into account (\ref{spec3}), we can write
$$
\Psi_{\alpha \beta}=e^{zx+\xi ({\bf t}, z)}\left (\sum_i
a_i^{\alpha}b_i^{\beta}\nu_{\beta}^{-1} +
z^{-1}\Bigl (S_{\alpha \beta}\! -\! \sum_i
a_i^{\alpha}b_i^{\beta}\zeta (x-x_i)\Bigr ) +O(z^{-2})\right ).
$$
Comparing with (\ref{m7}), we conclude that
\beq\label{spec5}
\sum_i a_i^{\alpha}b_i^{\beta}=\nu_{\alpha}\delta_{\alpha \beta}.
\eeq

It is easy to see that the Hamiltonian and the Lax matrix are invariant with respect to the
gauge transformation
\beq\label{spec6}
{\bf a}_i\longrightarrow W^{-1} {\bf a}_i, \qquad
{\bf b}_i^T\longrightarrow {\bf b}_i^T W
\eeq
with arbitrary non-degenerate $n\times n$ matrix $W$. Therefore, 
after the transformation $G\to W^{-1}gW$ the matrix $G$ can always
be regarded as diagonal matrix, as in (\ref{spec5}), with the eigenvalues being the same
as nonzero eigenvalues $\nu_{\alpha}$ of the $N\times N$ matrix $F$.

\section{Dynamics of poles in the higher times}

Our basic tool is equation (\ref{m11b}). Substituting $\Psi$, $\Psi^{\dag}$ in the form
(\ref{t3}), (\ref{t4}) and $\xi^{(1)}$ in the form (\ref{t3a}), we have:
\beq\label{ht1}
\begin{array}{l}
\displaystyle{
\frac{1}{2\pi i}\oint_{C_{\infty}} dz \, z^m \sum_{i,j}
a_i^{\alpha}c_i^{\nu}c_j^{*\nu}b_j^{\beta}\Phi (x-x_i, \lambda_{\nu})
\Phi (x-x_j, -\lambda_{\nu})}
\\ \\
\displaystyle{\phantom{aaaaaaaaaaaaaaa}
=\sum_i \p_{t_{\nu, m}}x_i a_i^{\alpha}b_i^{\beta}
\wp (x-x_i)+\sum_i \p_{t_{\nu , m}}(a_i^{\alpha}b_i^{\beta})\zeta (x-x_i)}
\end{array}
\eeq
(no summation over $\nu$ here!).
Equating the coefficients in front of the second order poles at $x=x_i$, we get the 
relation
\beq\label{ht2}
\p_{t_{\nu, m}}x_i=\res_{\infty} \Bigl (z^m c_i^{*\nu}c_i^{\nu}\Bigr )=
\res_{\infty} \Bigl (z^m {\sf c}^{*\nu}E_i {\sf c}^{\nu}\Bigr ),
\eeq
where $E_i$ is the diagonal $N\times N$ matrix with matrix elements
$(E_i)_{jk}=\delta_{ij}\delta_{ik}$ (again, no summation over $\nu$). 
Summing over $i$, we get
\beq\label{ht3}
\p_{t_{\nu, m}}\sum_i x_i=\res_{\infty} \Bigl (z^m {\sf c}^{*\nu}{\sf c}^{\nu}\Bigr )
\eeq
Comparing with equation (\ref{e8}), we conclude that
\beq\label{ht4}
{\sf c}^{*\alpha}{\sf c}^{\alpha}=-\nu_{\alpha}z^{-2}+
\sum_{m\geq 2}z^{-m-1}\p_{t_{\alpha, m}}\sum_i x_i=
-\lambda '_{\alpha}(z)
\eeq
(no summation over $\alpha$!).
Now, we note that $E_i=-\p_{p_i}L$ and continute (\ref{ht2}) as the following 
chain of equalities, using (\ref{t6}), (\ref{t7}), (\ref{ht4}):
$$
\p_{t_m}x_i=\sum_{\nu}\res_{\infty} \Bigl (z^m {\sf c}^{*\nu}E_i {\sf c}^{\nu}\Bigr )=-
\sum_{\nu}\res_{\infty} \Bigl (z^m {\sf c}^{*\nu}\p_{p_i}L(\lambda_{\nu}) {\sf c}^{\nu}\Bigr )
$$
$$
=-\sum_{\nu}\res_{\infty} \Bigl (z^m\p_{p_i}\Bigl (
{\sf c}^{*\nu}L(\lambda_{\nu}) {\sf c}^{\nu}\Bigr )\Bigr )+
\sum_{\nu}\res_{\infty} \Bigl (z^m (\p_{p_i}
{\sf c}^{*\nu})L(\lambda_{\nu}) {\sf c}^{\nu}\Bigr )+
\sum_{\nu}\res_{\infty} \Bigl (z^m 
{\sf c}^{*\nu}L(\lambda_{\nu}) \p_{p_i}{\sf c}^{\nu}\Bigr )
$$
$$
=-\sum_{\nu}\res_{\infty} \Bigl (z^m\p_{p_i}\Bigl (
{\sf c}^{*\nu}L(\lambda_{\nu}) {\sf c}^{\nu}\Bigr )\Bigr )+
\sum_{\nu}\res_{\infty} \Bigl (z^m (\p_{p_i}
{\sf c}^{*\nu})k_{\nu}{\sf c}^{\nu}\Bigr )+
\sum_{\nu}\res_{\infty} \Bigl (z^m 
{\sf c}^{*\nu}k_{\nu} \p_{p_i}{\sf c}^{\nu}\Bigr )
$$
$$
=-\sum_{\nu}\res_{\infty} \Bigl (z^m\p_{p_i}\Bigl (
{\sf c}^{*\nu}k_{\nu} {\sf c}^{\nu}\Bigr )\Bigr )+
\sum_{\nu}\res_{\infty} \Bigl (z^m k_{\nu}
\p_{p_i}({\sf c}^{*\nu} {\sf c}^{\nu})\Bigr )
$$
$$
=\sum_{\nu}\res_{\infty} \Bigl (z^m \lambda'_{\nu}(z)\p_{p_i}k_{\nu}\Bigr ).
$$
Regarding $z$ as an independent variable, we apply the same argument as in 
\cite{PZ21} to obtain
\beq\label{ht5}
\p_{t_m}x_i=-\sum_{\nu}\res_{\infty} \Bigl (z^m \p_{p_i}\lambda_{\nu}(z)\Bigr ).
\eeq
In this way we obtain the first half of the higher Hamiltonian equations for poles
\beq\label{ht6}
\p_{t_m}x_i =\frac{\p H_m}{\p p_i}
\eeq
with the Hamiltonian
\beq\label{ht7}
H_m=\sum_{\alpha=1}^n \res_{\infty} \Bigl (z^m\lambda_{\alpha}(z)\Bigr ).
\eeq

The second half of the Hamiltonian equations for poles can be obtained by
taking the $t_2$-derivative of (\ref{ht2}) and using (\ref{t6}), (\ref{t7}). 
In this way we obtain:
\beq\label{ht8}
\p_{t_{\nu, m}}\dot x_i=\res_{\infty}\Bigl (z^m {\sf c}^{*\nu}[E_i, M(\lambda_{\nu})]
{\sf c}^{\nu}\Bigr ).
\eeq
A straightforward verification shows that
\beq\label{ht9}
[E_i, M(\lambda )]=2\p_{x_i}L(\lambda ).
\eeq
Recalling also that $\dot x_i=2p_i$, we rewrite (\ref{ht8}) as
\beq\label{ht10}
\p_{t_{\nu, m}}p_i=\res_{\infty}\Bigl (z^m {\sf c}^{*\nu}\p_{x_i}L(\lambda_{\nu} )
{\sf c}^{\nu}\Bigr )
\eeq
(no summation over $\nu$!). With the relation (\ref{ht10}) at hand, one can repeat
the chain of equalities after equation (\ref{ht4}) with the change $\p_{p_i}\to \p_{x_i}$
to obtain
\beq\label{ht11}
\p_{t_m}p_i=\sum_{\nu}\res_{\infty} \Bigl (z^m \p_{x_i}\lambda_{\nu}(z)\Bigr ),
\eeq
so that
\beq\label{ht6a}
\p_{t_m}p_i =-\frac{\p H_m}{\p x_i}
\eeq
with the same Hamiltonian (\ref{ht7}). 

Let us make some comments on a more general case when the tau-function for elliptic
solutions has a slightly more general form
\beq\label{pt17}
\tau (x, {\bf t})=e^{Q(x, {\bf t})}\prod_{i=1}^N \sigma (x-x_i({\bf t})),
\eeq
where
\beq\label{pt18}
Q(x, {\bf t})=c(x+t_1)^2 +(x+t_1)\sum_{j\geq 2}a_j t_j +b(t_2, t_3, \ldots )
\eeq
with some constants $c$, $a_j$ and a function $b(t_2, t_3, \ldots )$. 
Repeating the arguments leading to (\ref{ht5}), one can see that now the
first equation in  (\ref{e7}) will be modified as
\beq\label{pt19}
k_{\beta}=z -\alpha (z)+\zeta (\lambda_{\beta}), \qquad
\alpha (z)=2cz^{-1}+\sum_{j\geq 2}\frac{a_j}{j}\, z^{-j}.
\eeq
Instead of (\ref{ht5}) we will have
\beq\label{pt20}
\p_{t_m}x_i=-\sum_{\nu}\res_{\infty} \Bigl (z^m \p_{p_i}\lambda_{\nu}(z)(1-\alpha '(z))\Bigr ),
\eeq
so the Hamiltonian for the $m$-th flow will be a linear combination of $H_m$ and $H_j$ with
$1\leq j<m$.

\section{Dynamics of spin variables in the higher times}

The Hamiltonian dynamics of spin variables in the higher times can be derived 
by analysis of first order poles in (\ref{ht1}). Equating coefficients in front of
first order poles, we get the relation
$$
\p_{t_{\nu , m}}(a_i^{\alpha}b_i^{\beta})=\res_{\infty}\left (
z^m \sum_{j\neq i} a_{i}^{\alpha}c_i^{\nu}c_j^{*\nu}b_j^{\beta}\Phi (x_i\! -\! x_j, -\lambda_{\nu})
+z^m\sum_{j\neq i} a_{j}^{\alpha}c_j^{\nu}c_i^{*\nu}b_i^{\beta}\Phi (x_i\! -\! x_j, \lambda_{\nu})
\right )
$$
which can be rewritten as
$$\hspace{-2cm}
a_i^{\alpha}\left [\p_{t_{\nu , m}}b_i^{\beta} +\res_{\infty}\Bigl (
z^m c_i^{\nu}\sum_{j\neq i}c_j^{*\nu}b_j^{\beta}
\Phi (x_j\! -\! x_i, \lambda_{\nu})\Bigr )\right ]
$$
$$\hspace{2cm}
+b_i^{\beta}\left [\p_{t_{\nu , m}}a_i^{\alpha} -\res_{\infty}\Bigl (
z^m c_i^{*\nu}\sum_{j\neq i}c_j^{\nu}a_j^{\alpha}
\Phi (x_i\! -\! x_j, \lambda_{\nu})\Bigr )\right ]=0.
$$
Now we notice that
\beq\label{spin1}
\begin{array}{l}
\displaystyle{
\frac{\p L_{jk}(\lambda )}{\p a_i^{\alpha}}=-\delta_{ik}(1-\delta_{jk})
b_j^{\alpha}\Phi (x_j-x_i, \lambda ),}
\\ \\
\displaystyle{
\frac{\p L_{jk}(\lambda )}{\p b_i^{\beta}}=-\delta_{ij}(1-\delta_{jk})
a_k^{\beta}\Phi (x_i-x_k, \lambda ),}
\end{array}
\eeq
so the equation above can be written as
\beq\label{spin2}
\begin{array}{l}
\displaystyle{a_i^{\alpha}\left [\p_{t_{\nu , m}}b_i^{\beta} -\res_{\infty}\Bigl (
z^m {\sf c}^{*\nu}\frac{\p L(\lambda_{\nu} )}{\p a_i^{\beta}}\, {\sf c}^{\nu}\Bigr )\right ]}
\displaystyle{
+b_i^{\beta}\left [\p_{t_{\nu , m}}a_i^{\alpha} +\res_{\infty}\Bigl (
z^m {\sf c}^{*\nu}\frac{\p L(\lambda_{\nu} )}{\p b_i^{\alpha}}\, {\sf c}^{\nu}\Bigr )\right ]=0}.
\end{array}
\eeq
Having this equation at hand, one can repeat
the chain of equalities after equation (\ref{ht4}) with the changes $\p_{p_i}\to 
\p / \p a_i^{\beta}$, $\p_{p_i}\to 
\p / \p b_i^{\alpha}$
to obtain
\beq\label{spin3}
a_i^{\alpha}P_i^{\beta}-b_i^{\beta}Q_i^{\alpha}=0,
\eeq
where
\beq\label{spin4}
P_i^{\beta}=-\p_{t_m}b_i^{\beta}+\sum_{\nu} \res_{\infty}\Bigl (
z^m \frac{\p}{\p a_i^{\beta}}\, \lambda_{\nu}(z)\Bigr )=-\p_{t_m}b_i^{\beta}-
\frac{\p H_m}{\p a_i^{\beta}},
\eeq
\beq\label{spin5}
Q_i^{\alpha}=\p_{t_m}a_i^{\alpha}+\sum_{\nu} \res_{\infty}\Bigl (
z^m \frac{\p}{\p b_i^{\alpha}}\, \lambda_{\nu}(z)\Bigr )=\p_{t_m}a_i^{\alpha}-
\frac{\p H_m}{\p b_i^{\alpha}}.
\eeq

It follows from (\ref{spin3}) that
$$
\frac{Q_i^{\alpha}}{a_i^{\alpha}}=\frac{P_i^{\beta}}{b_i^{\beta}}=\Lambda_i^{(m)},
$$
and the equations (\ref{spin4}), (\ref{spin5}) acquire the form
\beq\label{spin6}
\p_{t_m}a_i^{\alpha}=a_i^{\alpha}\Lambda_i^{(m)}+\frac{\p H_m}{\p b_i^{\alpha}},
\eeq
\beq\label{spin6a}
\p_{t_m}b_i^{\beta}=-b_i^{\beta}\Lambda_i^{(m)}-\frac{\p H_m}{\p a_i^{\beta}}.
\eeq
The gauge transformation $a_i^{\alpha}\to a_i^{\alpha} q_i^{(m)}$,
$b_i^{\alpha}\to b_i^{\alpha} (q_i^{(m)})^{-1}$ with
$\displaystyle{q_i^{(m)}=\exp \left (\int^{t_m}\Lambda_i^{(m)}dt\right )}$
eliminates the terms with $\Lambda_i^{(m)}$, so we can put $\Lambda_i^{(m)}=0$.
We obtain the Hamiltonian equations of motion for spin variables
in the higher times:
\beq\label{spin7}
\p_{t_m}a_i^{\alpha}=\frac{\p H_m}{\p b_i^{\alpha}}, \qquad
\p_{t_m}b_i^{\alpha}=-\frac{\p H_m}{\p a_i^{\alpha}}.
\eeq
with $H_m$ given by (\ref{ht7}).

\section{How to obtain the first two Hamiltonians}

In order to find the Hamiltonians, we need to expand the spectral curve near 
$\lambda =0$. Using the expansion (\ref{Phi1}), we represent the equation of the 
spectral curve as
\beq\label{H1}
\det \Bigl (zI+F\lambda^{-1} +Q +S\lambda +O(\lambda^2)\Bigr )=0,
\eeq
where the matrices $Q$, $S$ are
\beq\label{H2}
Q_{ij}=p_i\delta_{ij}+(1-\delta_{ij})F_{ij}\zeta (x_i-x_j),
\eeq
\beq\label{H3}
S_{ij}=\frac{1}{2}\, (1-\delta_{ij})\,F_{ij}\Bigl (\zeta^2(x_i-x_j)-\wp (x_i-x_j)\Bigr ).
\eeq
We set
\beq\label{H5}
z=-\frac{\omega}{\lambda},
\eeq
then the equation (\ref{H1}) acquires the form
\beq\label{H1a}
\det \Bigl (\omega I-F -Q\lambda -S\lambda^2 +O(\lambda^3)\Bigr )=0.
\eeq
This equation has $n$ roots $\omega_{\alpha}$ such that 
\beq\label{H4}
\omega_{\alpha}=\omega_{\alpha}(\lambda )=\nu_{\alpha}+\omega_1^{(\alpha )}\lambda +
\omega_2^{(\alpha )}\lambda^2 +O(\lambda^3)
\eeq
and $N-n$ roots which are $O(\lambda )$. These roots are eigenvalues of the matrix
$F +Q\lambda +S\lambda^2 +O(\lambda^3)$. Expressing $\lambda$ through $z$ from equation
(\ref{H5}) and expanding in powers of $z^{-1}$, we have
\beq\label{H6}
\lambda_{\alpha}=-\frac{\nu_{\alpha}}{z}+\nu_{\alpha}\omega_1^{(\alpha )}z^{-2}-
(\nu^2_{\alpha}\omega_2^{(\alpha )}+\nu_{\alpha}(\omega_1^{(\alpha )})^2)z^{-3}+O(z^{-4}).
\eeq
Then
\beq\label{H7}
\begin{array}{l}
\displaystyle{
H_1=-\sum_{\alpha}\nu_{\alpha}\omega_1^{(\alpha )}}
\\ \\
\displaystyle{
H_2=\sum_{\alpha}(\nu^2_{\alpha}\omega_2^{(\alpha )}+\nu_{\alpha}(\omega_1^{(\alpha )})^2)}.
\end{array}
\eeq
We regard the matrix $Q\lambda +S\lambda^2$ as a small variation of the matrix $F$.
The idea is to find the variation of the eigenvalues (the corrections
$\omega_1^{(\alpha )}\lambda +
\omega_2^{(\alpha )}\lambda^2$ in (\ref{H4})) using first two orders of the
perturbation theory. 

Let $\psi^{(j)}$ be a basis in the $N$-dimensional space and $\tilde \psi^{(j)}$
be the dual basis such that $(\tilde \psi^{(i)}\psi^{(j)})=0$ at $i\neq j$. We take first
$n$ vectors to be
$$
\psi^{(\alpha )}_i=b^{\alpha}_i , \qquad
\tilde \psi^{(\alpha )}_i=a^{\alpha}_i,
$$
then 
$$
(\tilde \psi^{(\alpha )}\psi^{(\beta )})=\sum_ia^{\alpha}_ib^{\beta}_i=
\nu_{\alpha}\delta_{\alpha \beta}, \quad \alpha , \beta =1, \ldots , n.
$$
These vectors are eigenvectors of the (non-perturbed) matrix $F$ with nonzero eigenvalues:
\beq\label{H8}
F\psi^{(\alpha )}=\nu_{\alpha}\psi^{(\alpha )}, \quad
\tilde \psi^{(\alpha )}F=\nu_{\alpha}\tilde \psi^{(\alpha )}.
\eeq
The other $N-n$ vectors are chosen to be orthonormal:
$$
(\tilde \psi^{(i)}\psi^{(j)})=\delta_{ij}, \quad i,j=n+1, \ldots , N.
$$

In the first order of the perturbation theory we have:
\beq\label{H9}
\omega_1^{(\alpha )}=
\frac{(\tilde \psi^{(\alpha )}Q\psi^{(\alpha )})}{(\tilde \psi^{(\alpha )}\psi^{(\alpha )})}.
\eeq
The next coefficient, $\omega_2^{(\alpha )}$, is obtained in the second order of the
perturbation theory as
\beq\label{H10}
\omega_2^{(\alpha )}=\frac{(\tilde \psi^{(\alpha )}S\psi^{(\alpha )})}{(\tilde \psi^{(\alpha )}\psi^{(\alpha )})}+
\sum_{j\neq \alpha}\frac{(\tilde \psi^{(\alpha )}Q\psi^{(j )})\,
(\tilde \psi^{(j )}Q\psi^{(\alpha )})}{(\tilde \psi^{(\alpha )}\psi^{(\alpha )})
(\tilde \psi^{(j )}\psi^{(j )})(\nu_{\alpha}-\nu_j)}.
\eeq
In the denominator of the last term $\nu_j =\nu_{\beta}$ at $j=\beta$, $\beta =1, \ldots , n$ and
$\nu_j=0$ at $j=n+1, \ldots , N$. 

Using these formulas, we have:
\beq\label{H11}
\sum_{\alpha}\nu_{\alpha}\omega_1^{(\alpha )}=\sum_{\alpha}
(\tilde \psi^{(\alpha )}Q\psi^{(\alpha )})=
\sum_{\alpha}\sum_{i,j}a_i^{\alpha}Q_{ij}b_j^{\alpha}=
\sum_{i,j}F_{ji}Q_{ij}=\mbox{tr}\, (FQ),
\eeq
$$
\sum_{\alpha}\Bigl (\nu^2_{\alpha}\omega_2^{(\alpha )}+\nu_{\alpha}(\omega_1^{(\alpha )})^2\Bigr )
=\sum_{\alpha \neq \beta}\frac{\nu_{\alpha}(\tilde \psi^{(\alpha )}Q\psi^{(\beta )})
(\tilde \psi^{(\beta )}Q\psi^{(\alpha )})}{\nu_{\beta} (\nu_{\alpha}-\nu_{\beta})}+
\sum_{j=n+1}^N \sum_{\alpha}(\tilde \psi^{(\alpha )}Q\psi^{(j )})
(\tilde \psi^{(j )}Q\psi^{(\alpha )})
$$
$$\phantom{aaaaaaaaaaa}+\sum_{\alpha}\nu_{\alpha}^{-1}
(\tilde \psi^{(\alpha )}Q\psi^{(\alpha )})^2 +\sum_{\alpha}\nu_{\alpha}
(\tilde \psi^{(\alpha )}S\psi^{(\alpha )})
$$
$$
=\sum_{\alpha , \beta}\nu_{\alpha}^{-1}(\tilde \psi^{(\alpha )}Q\psi^{(\beta )})
\tilde \psi^{(\beta )}Q\psi^{(\alpha )})+
\sum_{j=n+1}^N \sum_{\alpha}(\tilde \psi^{(\alpha )}Q\psi^{(j )})
(\tilde \psi^{(j )}Q\psi^{(\alpha )})
+\sum_{\alpha}\nu_{\alpha}
(\tilde \psi^{(\alpha )}S\psi^{(\alpha )})
$$
$$
=\sum_{ijkl}\Bigl (\sum_{\alpha}\nu_{\alpha}^{-1}a_i^{\alpha}b_l^{\alpha}+
\sum_{r=n+1}^N \tilde \psi^{(r)}_i\psi^{(r)}_l\Bigr )Q_{ij}F_{jk}Q_{kl}+
\sum_{\alpha}\nu_{\alpha}a_i^{\alpha}S_{ij}b_j^{\alpha}.
$$
But
$$
\sum_{\alpha}\nu_{\alpha}^{-1}a_i^{\alpha}b_l^{\alpha}+
\sum_{r=n+1}^N \tilde \psi^{(r)}_i\psi^{(r)}_l=\delta_{il}
$$
(the completeness relation),
and so finally we obtain
\beq\label{H12}
\sum_{\alpha}\Bigl (\nu^2_{\alpha}\omega_2^{(\alpha )}+
\nu_{\alpha}(\omega_1^{(\alpha )})^2\Bigr )=
\mbox{tr}\, (QFQ)+\mbox{tr}\, (FSF).
\eeq

From (\ref{H11}) we obtain
\beq\label{H13}
H_1=-\mbox{tr}\, (FQ)=-\sum_i p_i F_{ii}-\sum_{i\neq j}F_{ij}F_{ji}\zeta (x_i-x_j)=
-\sum_ip_i,
\eeq
which is indeed the first Hamiltonian. 
The calculation of (\ref{H12}) is more involved. We have, after some cancellations:
$$
\mbox{tr}\, (QFQ)=\sum_i p_i^2+
\sum_{k\neq i}\sum_{j\neq i}F_{ij}F_{jk}F_{ki}\zeta (x_i-x_j)\zeta (x_k-x_i),
$$
$$
\mbox{tr}\, (FSF)=\frac{1}{2}\sum_l \sum_{i\neq j}F_{li}F_{ij}F_{jl}
\Bigl (\zeta ^2(x_i-x_j)-\wp (x_i-x_j)\Bigr ).
$$
Therefore,
\beq\label{H14}
H_2=\sum_i p_i^2 -\sum_{i\neq j}F_{ij}F_{ji}\wp (x_i-x_j)+{\cal F},
\eeq
where
\beq\label{H15}
{\cal F}=\sum {}^{'}\! F_{ij}F_{jk}F_{ki}\zeta (x_i-x_j)\zeta (x_k-x_i)+
\frac{1}{2}\sum {}^{'}\! F_{ij}F_{jk}F_{ki}\Bigl (\zeta ^2(x_i-x_j)-\wp (x_i-x_j)\Bigr )=0.
\eeq
Here $\sum {}^{'}$ means summation over all distinct indices $ijk$. The proof of identity
(\ref{H15}) is given in Appendix A. To conclude, we have reproduced the correct Hamiltonians
$H_1$ and $H_2$ within our approach. 

\section{Rational and trigonometric limits}

In the rational limit $\omega , \omega ' \to \infty$, $\sigma (\lambda )=\lambda$,
$\Phi (x, \lambda )=(x^{-1}+\lambda^{-1})e^{-x/\lambda}$ and the equation of the spectral
curve becomes
\beq\label{r1}
\det \Bigl (zI-L_{\rm rat} +\lambda^{-1}F\Bigr )=0,
\eeq
where
\beq\label{r2}
(L_{\rm rat})_{ij}=-\delta_{ij}p_i -(1-\delta_{ij})\frac{b_i^{\nu}a_j^{\nu}}{x_i-x_j}
\eeq
is the Lax matrix of the spin generalization of the rational Calogero-Moser model.
Let us rewrite the equation of the spectral curve in the form
\beq\label{r3}
\det \left ( \lambda I+F \frac{1}{zI-L_{\rm rat}}\right )=0.
\eeq
Expanding the determinant, we have:
\beq\label{r4}
\lambda^N +\sum_{j=1}^{n}D_j(z) \lambda^{N-j}=0, \qquad D_1(z)=\mbox{tr}\,
\Bigl (F \frac{1}{zI-L_{\rm rat}} \Bigr ),
\eeq
where we took into account that rank of $F$ is equal to $n\leq N$. Let us note that the
functions $\lambda_{\alpha}(z)$ are different nonzero roots of equation (\ref{r4}) and
the sum of these roots is equal to $-D_1(z)$. Therefore, we can write
\beq\label{r5}
H_m=-\sum_{\nu}\res_{\infty}\Bigl (z^m\lambda_{\nu}(z)\Bigr )=
\res_{\infty}\Bigl (z^m \mbox{tr}\, \Bigl (
F \frac{1}{zI-L_{\rm rat}} \Bigr )\Bigr )=\mbox{tr}\Bigl (FL_{\rm rat}^m\Bigr ).
\eeq
It is straightforward to check the commutation relation
\beq\label{r6}
[X,L_{\rm rat}]=F-I, \qquad X=\mbox{diag} \, (x_1, \ldots , x_N).
\eeq
Substituting it into (\ref{r5}), we see that
\beq\label{r7}
H_m=\mbox{tr}L_{\rm rat}^m.
\eeq
This is the result of paper \cite{PZ18} obtained there by another method. 

We now pass to the trigonometric limit. We choose the
period of the trigonometric (or hyperbolic) functions to be $\pi i/\gamma$, where
$\gamma$ is some complex constant (real for hyperbolic functions and purely imaginary
for trigonometric functions).
The second period tends to infinity.
The Weierstrass functions in this limit become
$$
\sigma (x)=\gamma^{-1}e^{-\frac{1}{6}\, \gamma^2x^2}\sinh (\gamma x),
\quad
\zeta (x)=\gamma \coth (\gamma x)-\frac{1}{3}\, \gamma^2 x.
$$
The tau-function for trigonometric solutions is \cite{PZspintrig}
\beq\label{trig1}
\tau =\prod_{i=1}^N \Bigl (e^{2\gamma x}-e^{2\gamma x_i}\Bigr ),
\eeq
so we should consider
\beq\label{trig2}
\tau =\prod_{i=1}^N \sigma (x-x_i)e^{\frac{1}{6}\, \gamma^2
(x-x_i)^2 +\gamma (x+x_i)}.
\eeq
Similarly to the KP case \cite{PZ21},
equation (\ref{e7}) with this choice acquires the form
\beq\label{trig3}
k_{\beta}=z+\gamma \coth (\gamma \lambda_{\beta} ).
\eeq

The trigonometric limit of the function $\Phi (x, \lambda )$ is
$$
\Phi (x, \lambda )=\gamma \Bigl (\coth (\gamma x)+\coth (\gamma \lambda )\Bigr )
e^{-\gamma x \coth (\gamma \lambda )}.
$$
For further calculations it is convenient to pass to the variables 
\beq\label{trig4}
w_i=e^{2\gamma x_i}
\eeq
and introduce the diagonal matrix $W=\mbox{diag}\, (w_1, w_2, \ldots , w_N)$. 
In this notation, the equation of the spectral curve acquires the form
\beq\label{trig5}
\det \Bigl (W^{1/2}(zI-(L_{\rm trig}-\gamma I))W^{-1/2}+\gamma (\coth (\gamma \lambda )-1)F
\Bigr )=0,
\eeq
where $L_{\rm trig}$ is the Lax matrix of the spin Calogero-Moser model with matrix elements
\beq\label{trig6}
(L_{\rm trig})_{ij}=-p_i\delta_{ij}-\frac{(1-\delta_{ij})
\gamma F_{ij}}{\sinh (\gamma (x_i-x_j))}=
-p_i\delta_{ij}-2(1-\delta_{ij})\frac{\gamma w_i^{1/2}w_j^{1/2}F_{ij}}{w_i-w_j}.
\eeq
Some simple transformations allow one to bring the equation of the spectral curve 
to the form
\beq\label{trig7}
\det \left (\omega I+2\gamma W^{-1/2}FW^{1/2}\, \frac{1}{zI\! -\! 
(L_{\rm trig}-\gamma I)}\right )=0,
\qquad \omega =e^{2\gamma \lambda}-1.
\eeq
Expanding the determinant, we have:
\beq\label{trig8}
\omega^N +\sum_{j=1}^{n}K_j(z) \omega^{N-j}=0,
\eeq
where we took into account that rank of $F$ is equal to $n\leq N$. In particular,
$$
K_1=\mbox{tr}\, Y, \quad K_2=\frac{1}{2} (\mbox{tr}^2 Y-\mbox{tr}\, Y^2),
$$
where $Y$ is the matrix
$$
Y=2\gamma W^{-1/2}FW^{1/2}\, \frac{1}{zI\! -\! 
(L_{\rm trig}-\gamma I)}.
$$
The coefficients
$K_j$ are expressed through the 
elementary symmetric polynomials $e_j=e_j(\omega_1, \ldots , \omega_n)$
of nonzero roots $\omega_\nu=\omega_{\nu}(z)$ of this equation 
as $K_j=(-1)^j e_j(\omega_1, \ldots , \omega_n)$. Therefore,
$$
\sum_{\nu}\lambda_{\nu}(z)=\frac{1}{2\gamma}\sum_{\nu}\log (1+\omega_{\nu} (z))
=\frac{1}{2\gamma}\log \prod_{\nu} (1+\omega_{\nu} (z))
$$
$$
=\frac{1}{2\gamma}\log \Bigl (\sum_{j=0}^n e_j (\omega_1, \ldots , \omega_n)\Bigr )=
\frac{1}{2\gamma}\log \Bigl (\sum_{j=0}^n (-1)^jK_j \Bigr ).
$$
From this we conclude that
\beq\label{trig9}
\sum_{\nu}\lambda_{\nu}(z)=
\frac{1}{2\gamma}\, \log \det \left [I-2\gamma
W^{-1/2}FW^{1/2}
\frac{1}{zI-(L_{\rm trig}-\gamma I)}\right ].
\eeq
Starting from this point, one can literally repeat the corresponding calculation
from \cite{PZ21} with the change of the rank 1 matrix $E$ to the rank $n$ matrix $F$ and
using the easily proved relation
\beq\label{trig10}
[L_{\rm trig},W]=2\gamma (W^{1/2}FW^{1/2}-W).
\eeq
The result is
\beq\label{trig11}
\begin{array}{c}
\displaystyle{\sum_{\nu}\lambda_{\nu}(z) =\frac{1}{2\gamma}\, \mbox{tr} \Bigl 
(\log (I-z^{-1}(L_{\rm trig}+\gamma I))-
\log (I-z^{-1}(L_{\rm trig}-\gamma I))\Bigr )}
\\ \\
\displaystyle{=-\frac{1}{2\gamma}\, \mbox{tr} \sum_{m\geq 1}\frac{z^{-m}}{m}
\Bigl ((L_{\rm trig}+\gamma I)^m -(L_{\rm trig}-\gamma I)^m \Bigr )}
\end{array}
\eeq
and
\beq\label{trig12}
H_m=\frac{1}{2\gamma (m+1)}\, \mbox{tr} \Bigl ((L_{\rm trig}+\gamma I)^{m+1} -
(L_{\rm trig}-\gamma I)^{m-1} \Bigr )
\eeq
which agrees with the result of paper \cite{PZspintrig}.

\section*{Appendix A: Proof of identity (\ref{H15})}
\def\theequation{A\arabic{equation}}
\setcounter{equation}{0}

\addcontentsline{toc}{section}{\hspace{6mm}Appendix A}

Here we prove identity (\ref{H15}) ${\cal F}=0$, where
$$
{\cal F}=\sum {}^{'}\! F_{ij}F_{jk}F_{ki}\zeta (x_i-x_j)\zeta (x_k-x_i)+
\frac{1}{2}\sum {}^{'}\! F_{ij}F_{jk}F_{ki}\Bigl (\zeta ^2(x_i-x_j)-\wp (x_i-x_j)\Bigr ).
$$
and $\sum {}^{'}$ means summation over all distinct indices $ijk$. Using the behavior 
of the $\zeta$-function under shifts by periods
$$
\zeta (x+2\omega )=\zeta (x)+2\eta , \quad \zeta (x+2\omega ' )=\zeta (x)+2\eta ',
$$
one can see that ${\cal F}$ is a double-periodic function of any of $x_i$. Consider, 
for example, the shift of $x_1$ by $2\omega$. The terms in 
${\cal F}(x_1+2\omega )-{\cal F}(x_1)$ proportional to $\eta^2$ are:
$$
-(2\eta )^2\!\!\sum_{j\neq k\neq 1} \! F_{1j}F_{jk}F_{k1}+\frac{1}{2}\, (2\eta )^2
\!\!\sum_{j\neq k\neq 1} \!F_{1j}F_{jk}F_{k1}+\frac{1}{2}\, (2\eta )^2
\!\! \sum_{i\neq k\neq 1}\! F_{1k}F_{ki}F_{i1}=0.
$$
The terms proportional to $\eta$ are:
$$
-2\eta \!\!\sum_{j\neq k\neq 1} \! F_{1j}F_{jk}F_{k1}\zeta (x_1-x_k)-
2\eta \!\!\sum_{j\neq k\neq 1} \! F_{1j}F_{jk}F_{k1}\zeta (x_1-x_j)
$$
$$
+2\eta \!\!\sum_{j\neq k\neq 1} \! F_{j1}F_{1k}F_{kj}\zeta (x_j-x_k)+
2\eta \!\!\sum_{j\neq k\neq 1} \! F_{j1}F_{1k}F_{kj}\zeta (x_k-x_j)
$$
$$
+2\eta \!\!\sum_{j\neq k\neq 1} \! F_{k1}F_{1j}F_{jk}\zeta (x_1-x_j)
+2\eta \!\!\sum_{j\neq k\neq 1} \! F_{j1}F_{1k}F_{kj}\zeta (x_1-x_j)=0.
$$
Therefore, we see that ${\cal F}(x_1+2\omega )={\cal F}(x_1)$.
The double-periodicity in all other arguments is established in the same way. 

Next, the function ${\cal F}$ as a function of $x_1$ may have poles only at the points
$x_i$, $i=2, \ldots , N$. The second order poles cancel identically in the obvious way. 
We find the residue at the
simple pole at $x_1=x_2$ as follows:
$$
-\sum_{k\neq 1,2}F_{12}F_{2k}F_{k1}\zeta (x_1-x_k)-
\sum_{j\neq 1,2}F_{1j}F_{j2}F_{21}\zeta (x_1-x_j)
$$
$$
+\sum_{k\neq 1,2}F_{21}F_{1k}F_{k2}\zeta (x_1-x_k)
+\sum_{j\neq 1,2}F_{2j}F_{j1}F_{12}\zeta (x_1-x_j)=0.
$$
Vanishing of the residues in all other points and for all other variables 
can be proved in the same way. 
We see that the function ${\cal F}$ is a regular elliptic function and, therefore,
it must be a constant. To find this constant, we set $x_j=j\varepsilon$ 
and tend $\varepsilon$ to 0.
Thanks to the fact that $\zeta (x)=x^{-1}+O(x^3)$,
$\wp (x)=x^{-2}+O(x^2)$ as $x\to 0$, we find:
$$
{\cal F}=\frac{1}{\varepsilon^2}\sum {}^{'}\!\frac{F_{ij}F_{jk}F_{ki}}{(i-j)(k-i)}+O(\varepsilon^2)
$$
Making the cyclic changes of the summation variables $(ijk)\to (jki)$ and
$(ijk)\to (kij)$, we have:
$$
{\cal F}=\frac{1}{3\varepsilon^2}\sum {}^{'}\! F_{ij}F_{jk}F_{ki}
\left (\frac{1}{(i-j)(k-i)}+ \frac{1}{(j-k)(i-j)}+\frac{1}{(k-i)(j-k)}\right )+O(\varepsilon^2)
$$
$$
=\frac{1}{3\varepsilon^2}\sum {}^{'}\! F_{ij}F_{jk}F_{ki}\,
\frac{(j-k)+(k-i)+(i-j)}{(i-j)(j-k)(k-i)}+O(\varepsilon^2)=O(\varepsilon^2).
$$
Therefore, we conclude that ${\cal F}=0$ and the identity (\ref{H15}) is proved. 

\section*{Acknowledgments}
\addcontentsline{toc}{section}{\hspace{6mm}Acknowledgments}

The research of A.Z. has been funded within the framework of the
HSE University Basic Research Program and the Russian Academic Excellence Project '5-100'.

\end{document}